\documentclass[sigconf]{acmart}
\AtBeginDocument{%
  \providecommand\BibTeX{{%
    \normalfont B\kern-0.5em{\scshape i\kern-0.25em b}\kern-0.8em\TeX}}}

\setcopyright{none}
\acmYear{2021}
\acmDOI{10.1145/1122445.1122456}
\acmConference[FAccTRec2021]{4th FAccTRec Workshop: Responsible Recommendation}
\acmBooktitle{4th FAccTRec Workshop: Responsible Recommendation}
\acmPrice{15.00}
\acmISBN{978-1-4503-XXXX-X/18/06}

\usepackage[linesnumbered,ruled,vlined]{algorithm2e}
\SetKwInput{KwInput}{Input}                
\SetKwInput{KwOutput}{Output}  
\usepackage{caption}
\usepackage{subcaption}
\begin{document}

\title{An Adaptive Boosting Technique to Mitigate Popularity Bias in Recommender System}




\author{Ajay Gangwar}
\affiliation{%
 \institution{Indian Institute of Technology, Ropar}
 \country{India}
 }
\email{2019aim1002@iitrpr.ac.in}

\author{Shweta Jain}
\affiliation{%
  \institution{Indian Institute of Technology, Ropar}
 \country{India}
}
\email{shwetajain@iitrpr.ac.in}







\begin{abstract}
The observed ratings in most recommender systems are subjected to popularity bias and are thus not randomly missing. Due to this, only a few popular items are recommended, and a vast number of non-popular items are hardly recommended. Not suggesting the non-popular items lead to fewer products dominating the market and thus offering fewer opportunities for creativity and innovation. In the literature, several fair algorithms have been proposed which mainly focused on improving the accuracy of the recommendation system. However, a typical accuracy measure is biased towards popular items, i.e., it promotes better accuracy for popular items compared to non-popular items. This paper considers a metric that measures the popularity bias as the difference in error on popular items and non-popular items. Motivated by the fair boosting algorithm on classification, we propose an algorithm that reduces the popularity bias present in the data while maintaining accuracy within acceptable limits. The main idea of our algorithm is that it lifts the weights of the non-popular items, which are generally underrepresented in the data. With the help of comprehensive experiments on real-world datasets, we show that our proposed algorithm outperforms the existing algorithms on the proposed popularity bias metric.
\end{abstract}

\ccsdesc[500]{Information Filtering Systems~Recommender systems}
\keywords{Recommender systems, Popularity Bias, Fairness}
\maketitle
\section{Introduction}
Recommendation systems have become an essential part of our lives, from movies we watch on OTT platforms to shopping on e-commerce websites to the news we read on the internet. In this information age, where the information is available in abundance or huge volumes, recommendation systems make our lives simpler by filtering information for us according to our taste.

Recommender systems recommend relevant items to the users based on the user's previous data. The primary goal of recommendation engines is to identify the items that a particular customer might buy or be interested in based on the previous ratings. They do so by anticipating the ratings that users would give to the items, and the more data we have for a user, the more precisely we can predict their ratings. Netflix (movie recommendation system), Spotify (music recommender system), and Amazon (product recommendation system) are some examples of recommendation systems that we see in our daily lives.

Although recommender systems are quite popular, there are concerns regarding the fairness of these systems amongst the research community. Recommender systems primarily face fairness issues at two levels, user-level \cite{20,21}
and item level 
\cite{11,26}. This paper considers the most important fairness issue, namely the popularity bias at the item level. 
The items rated by most of the users or have received high ratings are known as popular items, and the items rated by very few users or not rated at all are known as non-popular items. The underlying notion behind popularity bias is that people are more inclined to offer comments on mainstream or popular products than on non-popular items. As a result, reported user response is skewed towards popular products rather than genuine user interest. Thus, frequently rated items or popular items receive a lot of exposure, but less popular or niche items are underrepresented in a recommendation. Recommending only popular items is problematic for two reasons: first, not everyone wants to follow the taste of the mainstream crowd, and second, it makes it harder for new goods to gain user attention.


One example of how just suggesting popular things affects companies is that a lawsuit was recently filed in the United States against Google for only showing advertising for popular items and not giving opportunities to less popular and newer items on the market \cite{23}. 
Because just a few items are displayed most of the time, this sort of activity generates a monopoly in the market, which is unsuitable for any firm. It will hamper the possibilities of creativity and innovation in the products or items.

Exploring and mitigating popularity bias in recommender systems is not a new problem and have been previously explored in many works
\cite{8,9,10,12,35}
. However, the existing works focus either on improving the accuracy of the overall recommendation system or exploring the non-popular items through diversification. Since more ratings are available for the popular items, the accuracy is inherently biased towards the popular items. A highly accurate recommender system may have very high accuracy on popular items but a very low accuracy on non-popular items. On the other hand, naive diversification may lead to poor accuracy of the overall recommender system.
In this work, we define popularity bias as the difference between errors on non-popular items and popular items. A fair recommender system should perform equally well on both popular and non-popular items thus ensuring a balanced accuracy of the recommender system. The main aim is to reduce popularity bias from the data and recommend relevant items to the user based on prior data while keeping the error as low as possible.
\subsection{Contributions}
The main contributions of this paper  are as follows:
\begin{itemize}
\item 
We provide a metric that quantitatively measures the popularity bias present in the data. The previous papers have used error as a metric that is inherently biased towards the popular items and thus may lead to poor accuracy on non-popular items.
\item To reduce the popularity bias, we propose a novel algorithm, namely FairBoost, that significantly reduces the popularity bias present in the data without deteriorating on the error.

\item We compare our proposed algorithm with the existing algorithms that claim to remove popularity bias and show that our algorithm outperforms the existing algorithms on the proposed popularity bias metric.
\end{itemize}

\section{Related Works}
Fairness in recommender systems has been the center of discussion for a long time since these systems suffer from different types of biases present in the data. Due to these biases, it can not represent the entire population, and due to this, the results produced by recommender systems are biased, so there is an utmost need to solve this problem. The most common types of biases are gender bias \cite{20,21}, racial bias\cite{27}, selection bias \cite{11,26}, exposure bias \cite{5,6}, position bias \cite{24,25}, and popularity bias \cite{8,9,10,12,35}. The focus of this paper is on popularity bias.

It is crucial to recommend non-popular items as they are the ones that are less likely to be discovered. \citet{8} have proposed a customized re-ranking diversification strategy that aims to boost non-popular items representation while retaining acceptable recommendation accuracy. The focus in \cite{8} was to achieve a trade-off between coverage of popular and non-popular items. To achieve this coverage, non-popular items that consumers could appreciate were randomly selected and presented to the user without being displayed in some order of preference. Thus the proposed method could result in poor accuracy for non-popular items.

To address the popularity bias, \citet{9} and \citet{10} employed the Inverse Propensity Score (IPS) method. Propensity-based techniques have previously been used in causal inference and observational studies but are used for the first time in the recommender system by \citet{10}. To apply the IPS approach, we need propensities, and the authors in \cite{10} have discussed primarily two propensity estimation models,  via Naive Bayes, and the other is via logistic regression. \citet{11} has also discussed a
straightforward approach to measure the propensities. However, the performance of these propensity-based methods is heavily influenced by the model used to estimate propensity. The algorithms require actual probabilities to function correctly which are difficult to estimate \cite{11}. As a result, a more effective strategy is required.


Furthermore, all of the above works focus on reducing recommender system errors without explicitly being fair to non-popular items.Two exceptions to this are \cite{36, 37}. \cite{36} mitigate the issue of popularity bias by providing a metric similar to the individual fairness metric used in machine learning. The idea is to equalise the outcomes across all individual items. On the other hand, \cite{37} focuses on group fairness by grouping the items together and equalizing the true positives across all groups. Our work focuses on group fairness, and instead of just equalizing true positives, we sought to equalize the total error across the groups.

Machine learning algorithms typically discriminate based on biased historical data. Data manipulation is used in certain preprocessing techniques to address this discrimination problem. Massaging \cite{15} , re-weighting \cite{16} , uniform or preferential sampling \cite{17} , and data augmentation \cite{18} are examples of preprocessing methods that might have an impact on data distribution. By altering instance labels, giving different weights, under or oversampling instances, and producing pseudo instances, they attempt to correct imbalances between protected and non-protected groups within the data. \citet{13} demonstrated Adafair, an Adaboost-based fairness-aware classifier that adjusts the weights of the instances in each subsequent round while explicitly addressing class imbalance by optimizing the number of ensemble models for balanced classification error. Adapting this idea, we examine the use of boosting algorithm in the recommender system. The idea is to increase the weights of non-popular items to maintain the balance between popular and non-popular items. We show that this technique can significantly reduce the popularity bias present in the system.

\section{The Model}
Let us consider the set of users ${U}$ and the set of items ${M}$, with the number of users $|U| = k$ and the number of items $|M|=l$. $A=U\times{M}$ represents the collection of (user, item) pairs with each entry $A_{u,m}$ denoting the true rating provided by user $u$ to item $m$. Integer values ranging from $1$ to $5$ can be used for representing each entry, with $1$ representing low interest and $5$ indicating strong interest. Because consumers only rate a small portion of the items, many ratings are typically unknown.
The main aim is to develop an algorithm that generates the best possible predicted rating matrix $\hat{A}$. In the predicted rating matrix $\hat{A}$, each entry $\hat{A}_{um}$ denotes the probable rating given by user $u$ to item $m$. To achieve the goal, ideal loss function to find predicted rating matrix is defined as: 
\begin{equation}
    L_{ideal}(\hat{A}) = \frac{ 1 }{ | A | }  \sum_{u=1}^k\sum_{m=1}^l { \delta_{um} ( A_{um} , \hat{ A }_{um} ) }
\end{equation}
where $\delta_{um}$ can be taken as mean squared error (MSE) i.e. $$\delta_{ um }^{ MSE } ( A ,\hat{ A } )=( A_{ u m } - \hat{ A }_{ u  m } )^{ 2 }$$  or minimum absolute error (MAE) or i.e. $$\delta_{ u  m }^{ MAE } ( A , \hat{ A } ) = | A_{ u  m } - \hat{ A }_{ u  m } |$$.

The ideal loss function cannot be calculated since the majority of the elements in the actual rating matrix are missing. Therefore, to build an effective recommender system, we need to estimate the ideal loss as accurately as possible using only the ratings present in the true rating matrix. The simplest straightforward estimator for estimating ideal loss, also known as a naïve estimator is used which is defined as follows:
\begin{equation}
    L_{naive}(\hat{A}) = \frac{\sum_{(u,m):B_{um}=1} \delta_{um}(A_{um},\hat{A}_{um})} {|\{(u,m):B_{um}=1\}|}
\end{equation}
where $B$ is an observed rating matrix and each entry $B_{um}=1$ if user $u$ has given rating to item $m$ otherwise $B_{um}=0$. 


The loss over observed ratings is calculated using naive loss, which is defined as the sum of loss overall user-item pairs in the data divided by the total number of user-item pairs in the true rating matrix. It is easy to see that when the data are missing fully at random, this naïve estimator is unbiased i.e.
$$E[L_{naive}(\hat{A})] = L_{ideal}(\hat{A})$$
However, when the data are not missing completely at random which happens in the presence of popularity bias where there is more rating available for popular items as opposed to that of non popular items, it is shown that the naive loss may not exactly correspond to the ideal loss function \cite{10,22} i.e.
\begin{equation}
    E[L_{naive}(\hat{A})] \neq L_{ideal}(\hat{A})
\end{equation}

One popular approach to mitigate the popularity bias is by using Inverse Propensity Score (IPS) approach \cite{9,10}. The main idea behind this approach is to build a pseudo missing completely at random dataset by weighting all the observed ratings by the inverse of their propensity score. 
It is easy to theoretically show that IPS loss is an unbiased estimator and hence can be used to remove popularity bias \cite{11}.

The IPS estimator's unbiasedness is desired; nevertheless, this feature is dependent on the true propensity scores which need to be approximated using various approaches.  These methods majorly suffers from two problems \cite{14}. One, the IPS estimator is no longer an unbiased estimator if a propensity estimation model is not stated appropriately. Another, the inverse of the propensities may be substantial, so the IPS estimator has a high variance problem. As a result, building learning methods that are resilient to misspecification of propensity and estimator variation is crucial for applying the approaches to real-world MNAR situations.

\citet{11} devised an asymmetric tri-training technique based on the asymmetric tri-training approach used in unsupervised domain adaptation to tackle the challenges that an IPS approach faced. It employs three rating predictors, two of which were used to create a pseudo rating dataset and the third of which was used to train the model on these pseudo ratings. The problem with this method is that the size of the dataset reduces after the algorithm is applied, making it impossible to accurately estimate the ratings of all the items. Because data in MNAR datasets are typically sparse, we must make appropriate use of it in order to predict ratings. 
Boosting technique on the other hand would be ideal because they utilise the majority of their data in different iteration stages, and one can also use upweighting to boost non-popular items in the same way that incorrectly categorised points are boosted in subsequent iterations.

\subsection{Quantifying popularity bias}
So far, the research community has been focusing on mitigating the popularity bias so as to improve the overall accuracy of the system. As mentioned in the Introduction section, resolving popularity bias as a standalone problem is highly required in order to prevent monopoly in the system. A typical accuracy measure is not a good metric for popularity bias because it is biased towards popular items, i.e., it promotes better accuracy for popular items compared to non-popular items. This is because more number of ratings are given to popular items. 

 One naive way to resolve popularity bias is to recommend non-popular items randomly thereby exploring these items. Another possibility to avoid the monopoly is through diversification \cite{33,2} which ensures that all group of items must be recommended some number of times.
 However, both naive or diversification methods could potentially lead to dissatisfaction amongst the users of the recommendation system. Thus, we need a metric that not only ensures the accuracy of the overall system, but also prevent the issue of popularity bias.   

We define the popularity biasedness of any algorithm by the difference in the error it achieves between the non-popular items and popular items. Let $\mathcal{PS}$ be the set of popular items and let $\mathcal{NPS}$ be the set of non-popular items. 
We employ the threshold $\tau$ to identify which of the items are popular and which are not.  The set of all those items that obtained more than the threshold number of ratings will be regarded popular, whereas the set of items that received less than the threshold number of ratings will be considered non-popular.$\tau$ is a dataset-specific parameter which can be taken as input. Then the popularity biasedness of the algorithm with predicted rating matrix $\hat{A}$ is defined as:
\begin{equation}\label{errd}
    \begin{split}
        PB(\hat{A},\tau) &=
        \frac{\sum_{(u,m) \colon m\in \mathcal{NPS}}\delta_{u,m}(A_{um},\hat{A}_{um}) B_{um}}{\sum_{(u,m) \colon m \in \mathcal{NPS}}B_{um}} - \\
        &\frac{\sum_{(u,m) \colon m \in \mathcal{PS}}\delta_{u,m}(A_{u,m},\hat{A}_{u,m}) B_{um}}{\sum_{(u,m) \colon m \in \mathcal{PS}}B_{um}}
    \end{split}
\end{equation}
where $\delta_{u,m}$ can be taken as MSE(mean squared error) or MAE(minimum absolute error).
This popularity bias metric separates the items into two groups: popular and non-popular. After that, the mean error is computed separately for non-popular and popular sets. Finally, the difference between them is used to determine the popularity bias $PB(\hat{A},\tau)$.
The goal is to minimize the $PB(\hat{A},\tau)$ to reduce the effect of popularity bias.

\subsection{Proposed Algorithm : FairBoost}
Boosting is a method of creating a powerful learner by combining several weak base learners. Adaboost\cite{19} is one such approach, which iteratively calls weak base learners after modifying the weights based on misclassified data points in each iteration. Because it separates the learning problem into numerous sub-problems and then combines their answers into an overall model, we believe boosting is a good fit for our problem. In sub-models, the popularity bias problem is easier to address than in the entire complicated model. As popular items frequently appear in the recommender system's results, the weights of the non-popular items must be increased in a way such that it shows  up in the recommender system's results. In the classification setting, the AdaBoost algorithm internally boosts the weights of erroneously categorised data points. The way we increase the weights of incorrectly classified data points in Adaboost, we increase the weights of non-popular items while keeping the accuracy on these items in mind. Finally, we adjust the reweighting procedure of the AdaBoost algorithm to make it more fair and obtain the FairBoost Algorithm.
Adaboost is not a novel technique for recommender systems; it was previously used to improve accuracy when using a decision tree as the base learner \cite{34}. However, we are the first to use and modify this notion to reduce popularity bias in recommender systems.


Algorithm 1 depicts the training phase of FairBoost. To establish its significance in the training dataset, FairBoost also provides weight to each training example. When the given weights are high, that set of training user-item pairs is more likely to influence the training set. Similarly, user-item pairs with low weights will have a little impact on the training dataset. At first, all of the user-item pairs will be given same weight of $1/n$, where $n$ is the number of user item pairs.
Fairboost additionally employs a popularity bias-related cost denoted by $cost_{um}$ for each user-item pair $(u, m)$, which try to maintain similarity and reduce the popularity bias that exists between popular and non-popular sets for current learners. 
We initialize $cost_{um}$ to zero for all $(u,m)$. Then the user item pairs are sampled using sample weights $w_{um}$, and a weak learner $\hat{A}^{j}$ is trained on these sampled points. Let $S_j$ is the set of sampled user-item pairs at $j^{th}$ iteration. The error rate $err_{j}$ is computed using:
\begin{equation}
    err_{j} = \sum_{(u,m) \in S_j} w_{um} 
    \left (1- e^{\frac{-(A_{um}-\hat{A}^j_{um})}{\max(A_{um}-\hat{A}^j_{um})}} \right )
\end{equation}
where $w_{um}$ is the weight of $(u,m)$ user-item pair in the sampled user-item pairs set $S_j$, $A_{um}$ is the actual rating of user-item pair and $\hat{A}^j_{um}$ is the rating predicted by the current base learner for user-item pair $(u,m)$.
Following that, we use the following formula to calculate the the weight $\alpha_{j}$ depicting the influence of the base learner $j$ in predicting ratings. 
\begin{equation}
    \alpha_{j}= \frac{1}{2} \log{\left (\frac{1-err_{j}}{err{j}} \right)}
\end{equation}
After that popularity bias ${PB( \hat{A}^j ,\tau)}$ is computed for the current base learner using Equation (\ref{errd}). 

Next popularity bias related cost, $cost_{um}$ is computed for all the user-item pairs $(u,m)$ in the sampled user-item pairs. Since we conduct random sampling every time, popular items become non-popular in some iterations and non-popular items become popular in other iterations.
We seek to achieve similarity between popular and non-popular sets or eliminate popularity bias. As a result, in the current iteration, a popularity bias-related cost is employed to preserve this similarity or reduce popularity bias. $cost_{um}$ is given by :
\begin{equation}
        cost_{um}=
        \begin{cases}
            |{PB( \hat{A}^j ,\tau)}|,if\big[(A_{um}-\hat{A}^{j}_{um}) > \epsilon_{1} AND\\ 
            \mspace{28mu} | {PB( \hat{A}^j ,\tau)}| > \epsilon_{2}
            ,m \in \mathcal{PS}, {PB( \hat{A}^j ,\tau)} > 0 \big]\\
            | {PB( \hat{A}^j ,\tau)}|,if\big[(A_{um}-\hat{A}^j_{um}) > \epsilon_{1} AND \\
            \mspace{28mu} | {PB( \hat{A}^j ,\tau)}| > \epsilon_{2} ,m \in \mathcal{NPS}, {PB(\hat{A}^j ,\tau)} < 0 \big]\\
            0,     otherwise
        \end{cases}
\end{equation}
It should be noted that $\epsilon_1$ is a hyperparameter used to set a bound on the difference between true and predicted ratings, and $\epsilon_2$ is a hyperparameter used to set a bound on popularity bias. Both of these parameters must be tuned for the algorithm to work properly.
For all user-item pairs, if item belongs to popular set, 
popularity bias ${PB( \hat{A}^j ,\tau)}$ is greater than zero and if difference between true and predicted rating is greater than $\epsilon_{1}$ and absolute value of popularity bias for current base learner is greater than $\epsilon_{2}$, then $cost_{um}$ assigned to the pair is $ | {PB( \hat{A}^j ,\tau)}|$. Here, a popular item has become unpopular, therefore we are weighing it more. For all user-item pairs, if item belongs to non-popular set, popularity bias 
${PB( \hat{A}^j ,\tau)}$ is less than zero and if difference between true and predicted rating is greater than $\epsilon_{1}$ and absolute value of popularity bias for current base learner is greater than $\epsilon_{2}$, then  $cost_{um}$ assigned to the pair is $ | {PB( \hat{A}^j ,\tau)}|$. We are upweighting this item because it is already unpopular. And for all other user-item pairs $cost_{um}$ of zero is assigned.
Then the weights are updated for the next round after computing costs using :
\begin{equation}
    w_{um} \leftarrow \frac{1}{Z_{j}}w_{um}.e^{\alpha_{j}.(A_{um}-\hat{A}^j_{um})}.(1+cost_{um})
\end{equation}
where $Z_{j}$ is a factor used for normalizing weights. The user-item pairs having the larger error are given greater weight, so they can be predicted accurately in the following iteration. Weights are updated for all sampled user-item pairs by multiplying $w_{um}$ with exponential of product of current base learner weight and difference between true and predicted rating so that the examples that are having more error gets more weight in next round and the examples having less errors gets less weight in next round. Weights are also multiplied by (1+$cost_{um}$). It is done for all those examples that were treated unfairly during current round.

And once the number of rounds is reached, the algorithm converges.
The algorithm will generate a number of base learners equal to the number of rounds, which will then be merged using the weights $\alpha_j$ or the amount of influence they have to produce the estimator $\hat{A}^j$ as an output. Now, in order to forecast the rating of a new user-item pair, it will go through all of the base learners. The predicted ratings from all these learners are weighted with the corresponding weights of the base learners and then combined to provide the expected rating for the new user-item pair.


\begin{algorithm}[h]
\caption{FairBoost Algorithm to mitigate popularity bias}
\label{Algorithm:algo}
\KwInput{A=$(X_{um},Y_{um})^{N},M,\epsilon_{1},\epsilon_{2}$}
\KwOutput{ Estimator $B$}
Initialize $w_{um}$ = $\frac{1}{N}$ and $cost_{um}=0$,for all user-item pair $(u,m) \in A$\\
\For{$j\gets1$ \KwTo $M$}{
    a) Weak learner $\hat{A}^{j}$ is trained using weights $w_{um}$ on training data \\
    b) Error rate $err_{j}$ is computed\\
    c) Weight is computed for the weak learner, $\alpha_{j} = \frac{1}{2}\log{\left (\frac{1-err_{j}}{err{j}} \right)}$\\
    d) Popularity Bias $ {PB( \hat{A}^j ,\tau)}$ is computed\\
    e) $cost_{um}$ related to popularity bias is computed.\\
    f) Distribution is updated as $w_{um} \leftarrow \frac{1}{Z_{j}}w_{um}.e^{\alpha_{j}.(A_{um}-\hat{A}^j_{um})}.(1+cost_{um})$, where $Z_{j}$ is a factor used for normalizing weights.
}
$B(x) = \sum_{j=1}^{M}\alpha_{j}\hat{A}^{j}(x)$
\end{algorithm}

\section{Experimental Analysis}
\subsection{Experimental setup}
\subsubsection{Datasets}
To show the effectiveness of our proposed algorithm, we have used the following realworld datasets. In all the datasets, training set and testing set have been created by considering 80\% and 20\% ratings respectively.
\begin{itemize}
    \item Netflix dataset\cite{29}
    The Netflix dataset consists of about 100 million MNAR five-star movie ratings (missing not at random). There are 480189 users and 17770 movies involved. Due to computational restrictions, we took 10 million of the most recent ratings by sorting them according to time. 
    There are 370811 users and 1962 movies in the training set, whereas there are 258603 users and 1962 movies in the test set.
    \item Yahoo dataset\cite{31}\\
    This dataset captures the preferences of the Yahoo! Music community and comprises almost 717 million ratings on 136 thousand songs provided by 1.8 million users. The data was gathered between 2002 and 2006. The ratings are on a scale of 1 to 5, with 1 being the lowest and 5 being the highest. We have sampled around 10 Million ratings from this set. Training set consists of 23179 users and 136737 songs while test set consists of 23179 users and 63261 movies.
    \item Amazon dataset\cite{32} \\
    This is an Amazon Movies and TV dataset which consists of around 4.6 million ratings. The ratings are on a scale of 1 to 5, with 1 being the lowest and 5 being the highest. There are 2088620 users and 200941 items involved.
    There are 1666901 users and 188083 items in the training set, whereas there are 562187 users and 80112 items in the test set.
    
    \item Movielens dataset\cite{30}\\
    This dataset is made up of 100K five-star movie ratings(missing not at random) gathered from a movie recommendation service . There are 1682 movies and 943 people involved. The data has been sorted by date. The dataset is divided into a training set and a test set, with the training set consisting of the previous 80\% of user-item pairs and the test set consisting of the rest or most recent 20\% of user-item pairs. There are 751 users and 1616 movies in the training set, whereas there are 301 users and 1448 movies in the test set.
\end{itemize}

\subsubsection{Compared methods}
We conducted comprehensive testing on the above mentioned datasets to demonstrate that our proposed algorithm lowers the system's popularity bias. Three algorithms were tested and their results were compared. We compared the Fairboost algorithm to that of Matrix Factorization with Inverse Propensity scoring \cite{10} and Asymmetric tri-training \cite{11}. The previous papers used error as a measure and focused on lowering the error rather than explicitly addressing the system's popularity bias. We compared several methods using the proposed popularity bias metric in Equation (~\ref{errd}). We keep the value of $\tau$ to be $100$ which means if an item received rating from more than $100$ users, then we call that item as popular item.

\subsubsection{Hyperparameter Tuning}
To adjust the parameters $\epsilon_{1}$ and $\epsilon_{2}$, we used a random search cv hyperparameter tuning procedure. Both parameters were tuned in the range $[10^{-5},1]$.

\subsection{Results and Discussion}
\begin{table*}[ht]
\footnotesize
    \renewcommand{\arraystretch}{1.5}
    \begin{tabular}{c|l|cccc}
    \hline
    \textbf{Datasets} & \textbf{Algorithm} & \textbf{Error} &\textbf{Error on popular items} & \textbf{Error on non-popular items} & \textbf{Popularity Bias} \\ 
    \hline
    {Netflix} & Matrix factorization & 1.0896 & 1.0675 & 1.1315 & 0.0639\\
    &Matrix factorization with IPS \cite{10}  & 1.0798  & 1.236  & 1.297   & 0.0610  \\ 
    & Asymmetric tri-training \cite{11}  &  1.2165     & 1.1076 & 1.1585  & 0.0508  \\
    &Adaboost \cite{34} & 1.1221 & 1.1087 & 1.1480 & 0.0393\\
    & \textbf{FairBoost} & 1.1481 & 1.1373 & 1.168 & \textbf{0.0315} \\
    \hline
    {Movielens} 
    &Matrix factorization & 1.0268 & 1.0154 & 1.0272 & 0.0117\\
    & Matrix factorization with IPS \cite{10} &1.0199 & 1.1630 & 1.1728 & 0.0098  \\
    & Asymmetric tr-training \cite{11} &1.1452 & 1.1519 & 1.1450 & 0.0069 \\
    &Adaboost \cite{34} & 1.10544 & 1.1011 & 1.1055 & 0.0080\\
    & \textbf{FairBoost} &1.0577 &1.0478 & 1.0511 & \textbf{0.0025}\\
    \hline
    {Amazon} 
    &Matrix factorization & 1.1468 & 1.2201 & 1.1223 & 0.0977\\
    & Matrix factorization with IPS \cite{10} & 1.1432 & 1.2264  & 1.3140 &  0.0876 \\
    & Asymmetric tr-training \cite{11} &1.1509 & 1.2126 & 1.1304 & 0.0822 \\
    &Adaboost \cite{34} & 1.1581 & 1.1821 & 1.1502 & 0.0319\\
    & \textbf{FairBoost} &1.2984 & 1.3286 & 1.3225  & \textbf{0.0119}\\ 
    \hline
    {Yahoo}
    &Matrix factorization & 1.4729 & 1.5436 & 1.4700 & 0.0736\\
    & Matrix factorization with IPS \cite{10} &1.3969 & 1.5876 & 1.5203 & 0.0673 \\
    & Asymmetric tr-training \cite{11} &1.5693 & 1.6111 & 1.5676 & 0.0434\\
    &Adaboost \cite{34} & 1.5712 & 1.6107 & 1.5696 & 0.0411\\
    & \textbf{FairBoost} &1.5021 & 1.5749 & 1.4991 & \textbf{0.0327}\\
    \hline
    \end{tabular}
    \\
    \caption{Results}
    \label{tab:results}
\end{table*}

The results from all the datasets are summarised in Table ~\ref{tab:results} which displays the results of 10 iterations of all the algorithms. Table~\ref{tab:results} can be used to make the following observations. The popularity bias
was reduced when IPS was used in conjunction with matrix factorization. However, the error on both popular and non-popular items had increased when compared to the baseline matrix factorization algorithm on all the datasets. Another finding was that the propensity estimation model chosen had a significant impact on the performance of propensity-based unbiased estimation approaches. We have used naive Bayes propensity estimation method for the comparison as it gave us the least value of popularity bias. The other methods such as user-item propensity gave us a good accuracy, however, the popularity bias was quite high. Another observation is that the popularity bias was also reduced when asymmetric tri-training was used when compared to matrix factorization based methods, as shown in table~\ref{tab:results}.

\begin{figure*}[htbp]
\begin{subfigure}{0.23\textwidth}
\includegraphics[width=\linewidth]{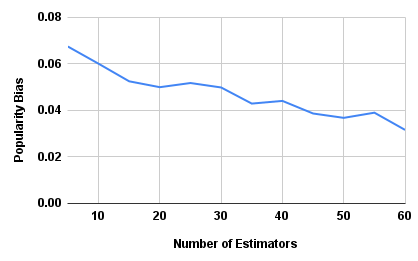}
\caption{Netflix dataset}
\label{error_difference_netflix}
\end{subfigure}
\begin{subfigure}{0.23\textwidth}
\includegraphics[width=\linewidth]{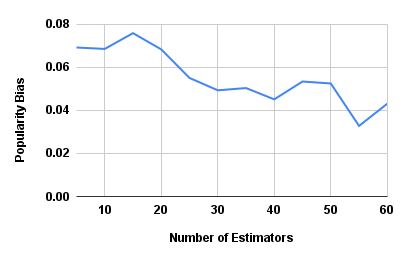}
\caption{Yahoo dataset}
\label{errd_yahoo}
\end{subfigure}
\begin{subfigure}{0.23\textwidth}
\includegraphics[width=\linewidth]{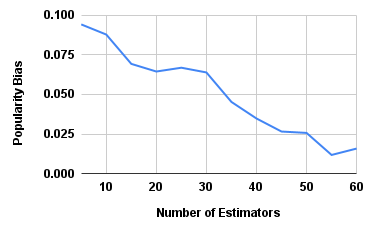}
\caption{Amazon dataset}
\label{errd_amzn}
\end{subfigure}
\begin{subfigure}{0.24\textwidth}
\includegraphics[width=\linewidth]{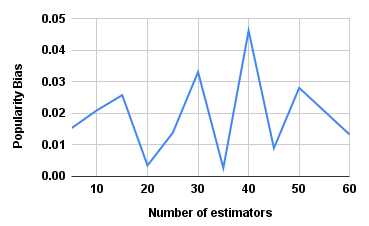}
\caption{Movielens dataset}
\label{error_difference_ml}
\end{subfigure}
\caption{Number of estimators vs Popularity Bias}
\label{fig1}
\end{figure*}
\begin{figure*}[htbp]
\begin{subfigure}{0.24\textwidth}
\includegraphics[width=\linewidth]{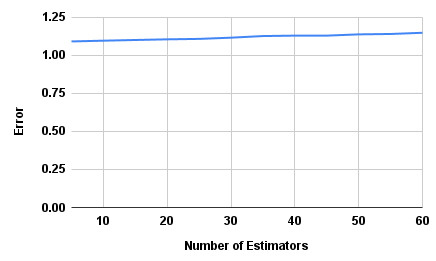}
\caption{Netflix dataset}
\label{error_netflix}
\end{subfigure}
\begin{subfigure}{0.24\textwidth}
\includegraphics[width=\linewidth]{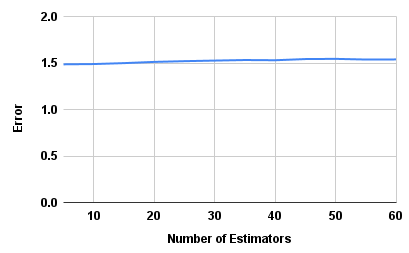}
\caption{Yahoo dataset}
\label{err_yahoo}
\end{subfigure}
\begin{subfigure}{0.24\textwidth}
\includegraphics[width=\linewidth]{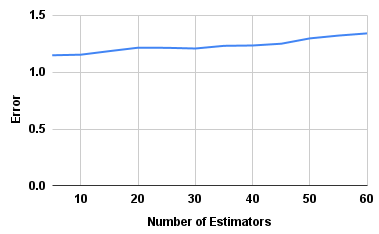}
\caption{Amazon dataset}
\label{err_amzn}
\end{subfigure}
\begin{subfigure}{0.24\textwidth}
\includegraphics[width=\linewidth]{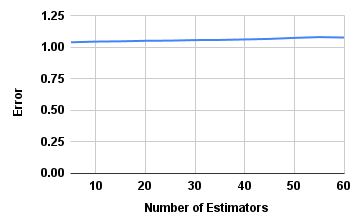}
\caption{Movielens dataset}
\label{error_ml}
\end{subfigure}
\caption{Number of estimators vs Error Plot}
\label{fig2}
\end{figure*}
As popular items appear frequently in the recommender system's results. So, in order for non-popular items to show up in the recommender system's results, their weightage must be increased in some way. 
In the classification setting, the AdaBoost algorithm internally boosts the weightage of wrongly categorised data points, so we thought we'd give that a try in ours because we also want to give non-popular items more weightage. 
Table~\ref{tab:results} shows that after applying Adaboost, the popularity bias has decreased on almost all of the datasets as compared to previously implemented methods. Inspired by how well Adaboost performed, we modified it to create the FairBoost algorithm (Algorithm ~\ref{Algorithm:algo}). When running the FairBoost algorithm, we employed matrix factorization as our underlying base learner. On all the datasets, Table~\ref{tab:results} shows that our proposed algorithm Fairboost significantly reduced the popularity bias when compared to other algorithms. In the following steps, the FairBoost algorithm tries to make non-popular items popular. We increase the weights of non-popular items in our algorithm in the following steps to make them popular, similar to how Adaboost increases the weights of inaccurately classified data points. Figures~\ref{error_difference_netflix},~\ref{errd_yahoo} and~\ref{errd_amzn} show that the popularity bias decreases as the number of estimators increases. This is because the FairBoost algorithm inherently gives more weight to non-popular items, causing the graph to decrease.  In the case of the movielens dataset, we got a zig-zag graph, as shown in Fig~\ref{error_difference_ml}. This is because the data is nearly fair; as the table~\ref{tab:results} shows, the error difference is extremely small, and ratings are also uniformly distributed. Figures ~\ref{error_netflix},~\ref{err_yahoo},~\ref{err_amzn}  and ~\ref{error_ml} show that the overall inaccuracy increases slightly. This is due to the fact that in succeeding steps, popular items weights are given less weight. As a result, FairBoost Algorithm minimises the popularity bias in each subsequent step while keeping the error increase within acceptable bounds.

\section{Conclusion and Future Work}
Adequate coverage of non-popular or long-tail items is critical to any business's success. Because almost all users are familiar with popular items, a recommender system's ability to recommend non-popular items will determine how well it introduces users to new experiences and products; however, it is well known that recommender systems are biased towards popular items.
\\
This paper proposed and compared FairBoost, an adaptive boosting algorithm, to previously implemented approaches for mitigating popularity bias in recommender systems. We also proposed a new metric to quantify popularity bias, because the error metric was insufficient for this purpose. On the four datasets, we were able to demonstrate that the FairBoost algorithm significantly reduces the popularity bias compared to other algorithms while keeping the error as low as possible.
One exciting area for future study would be to test this algorithm on other sorts of biases that might exist in the system such as selection bias, gender bias, and so on. Another thing we can try is to test the algorithm with different base learners.

\newpage
\bibliographystyle{ACM-Reference-Format}
\bibliography{ref}

\end{document}